\documentclass[epj,final]{svjour}
\usepackage{cite}
\usepackage[T1]{fontenc}
\usepackage[utf8]{inputenc}
\usepackage{amsmath} 
\usepackage{amssymb}
\usepackage{booktabs}
\usepackage{textcomp}
\usepackage{multirow}
\usepackage{subfig}
\usepackage{graphicx}
\usepackage{stackengine}
\usepackage{xcolor}
\usepackage{hyperref}
\hypersetup{
    colorlinks=true,
    citecolor=blue,
    linkcolor=blue,
    filecolor=magenta, 
    urlcolor=blue,
}
\usepackage{background}
\SetBgContents{As part of the Springer Nature SharedIt initiative, a full-text view-only version of the paper can be accessed \href{https://rdcu.be/bRg2t}{clicking here} }
\SetBgScale{1}
\SetBgColor{black}
\SetBgAngle{0}
\SetBgOpacity{1}
\SetBgPosition{current page.north east}
\SetBgVshift{-1cm}
\SetBgHshift{-10.5cm}

\providecommand{\tabularnewline}{\\}

\title{On the growth of non-motile bacteria colonies: an agent-based model
for pattern formation}

\author{Lautaro Vassallo\inst{1} \and David Hansmann\inst{1} \and Lidia A. Braunstein\inst{1,2}}

\institute{                    
  \inst{1} Departamento de Física, Facultad de Ciencias Exactas y Naturales, Universidad Nacional de Mar del Plata, and Instituto de Investigaciones Físicas de Mar del Plata (IFIMAR-CONICET), Deán Funes 3350, 7600 Mar del Plata, Argentina\\
  \inst{2} Physics Department and Center for Polymer Studies, Boston University, Boston, Massachusetts 02215, USA
}

\titlerunning{On the growth of non-motile bacteria colonies}
\authorrunning{Lautaro Vassallo et al}

\abstract{
In the growth of bacterial colonies, a great variety of complex patterns are observed in experiments, depending on external conditions and the bacterial species. Typically, existing models employ systems of reaction-diffusion equations or consist of growth processes based on rules, and are limited to a discrete lattice. In contrast, the two-dimensional model proposed here is an off-lattice simulation, where bacteria are modelled as rigid circles and nutrients are point-like, Brownian particles. Varying the nutrient diffusion and concentration, we simulate a wide range of morphologies compatible with experimental observations, from round and compact to extremely branched patterns. A scaling relationship is found between the number of cells in the interface and the total number of cells, with two characteristic regimes. These regimes correspond to the compact and branched patterns, which are exhibited for sufficiently small and large colonies, respectively. In addition, we characterise the screening effect observed in the structures by analysing the multifractal properties of the growth probability.}

\begin{document}
\maketitle
\sloppy

\section{Introduction}
The concept of active matter is relatively new within soft matter
physics; the fundamental units of this type of matter, called active
agents, have the particularity of absorbing energy from their environment
and dissipating it in order to move, grow or replicate, among other
activities \cite{DeMagistris2015}. Most of the examples of active
matter are biological, such as bacteria. Although they can be seen as
the simplest living organisms, they present interesting behaviours,
both individually and collectively.

Bacteria exhibit many different types of movement, depending on the
species and the environment, which determines the macroscopic appearance
of the colony. According to Henrichsen \cite{Henrichsen1972},
six types of motility can be identified: swimming, swarming, twitching,
darting, gliding (newer studies subdivide this category \cite{MarkJ.McBride2001}) and sliding. 
We focus on the last one, which is a
mechanism produced by the expansive forces of the colony, in combination
with special properties of the cell membranes characterised by low friction
with the substrate on which they grow; the bacteria do not move by
their own motors, but push each other by duplicating themselves and
competing for the same spaces. Despite its simplicity, its importance has been pointed out in more complex bacterial processes, such as in the formation, dispersion and restructuring of biofilms \cite{Mazza2016}. Although we will not strictly enter the biofilms field in this work, it is important to mention they are the focus of numerous studies, due to the complexity of the processes involved, their resistance to hostile environments, and the challenge they present to medicine \cite{Costerton1995, Mazza2016, Hall-Stoodley2004, OToole2000, Flemming2010}.

At the end of the 80s, Matsuyama \cite{Matsuyama1989},
Fujikawa and Matsushita \cite{Matsushita1990}, showed that the patterns
of bacterial colonies obtained in the laboratory could be fractal objects.
The properties of their patterns depend on two main factors: the concentration
of nutrients, which influences the growth rate of the colony, and
the concentration of agar, which determines the hardness of the substrate,
and therefore, the mobility of the bacteria. In the absence of special
forms of motility, the patterns were classified in a two-dimensional
phase diagram in which five characteristic patterns were identified:
diffusion-limited aggregation-like (DLA-like), Eden-like, dense branching
morphology (DBM), concentric ring and homogeneous disk-like. The experiments
were performed mainly with the species Bacillus subtilis \cite{Matsushita1990, Matsushita1998, Ohgiwari1992, Fujikawa1992}. Without self-propulsion, only DLA and Eden-like patterns are expected. 

At the theoretical level, continuous models are the most traditional
and extended way of studying the patterns exhibited by bacteria colonies.
In them, both bacteria and nutrients (or any other variable of interest)
are represented by density functions per unit area, and the spatio-temporal
evolution of the system is described by systems of reaction-diffusion
equations \cite{Matsushita1998, Kawasaki1997, Mimura2000, Giverso2015,
Golding1998,Matsushita1999,Kozlovsky1999,Matsushita2004,Lacasta1999,Tronnolone2018,Marrocco2010}.
These models are successful in describing a wide range of patterns, although
they are valid only at a mesoscopic scale. To represent growth at the microscopic
level, microorganisms must be represented by discrete mobile entities
(agents) \cite{Ben-Jacob1994,Farrell2013,Li1995,Melke2010}. In this scheme, on-lattice \cite{Li1995} and off-lattice \cite{Ben-Jacob1994,Farrell2013,Melke2010,Santalla} approaches can be chosen.

The motivation of this work is to propose a microscopic model that
can explain the experimental observations, based on the fact that sliding is dominated by the mechanical interaction between the bacterial cells. As has been said, continuous models work well only on a mesoscopic scale, whereas in agent-based models, if space is discretised as Euclidean networks, mechanical laws cannot be used. In spite of being computationally expensive, in this work, we choose an off-lattice model, in order to represent our agents as rigid bodies governed by
laws of mechanics. Thus, we can analyse the growth of bacterial colonies on a microscopic level. The off-lattice approach also avoids anisotropies in the patterns exhibited by the colonies induced by the discretization of the space. 

A typical way to characterise the complex structures that arise in
surface growth is by means of the Hausdorff dimension, often referred to as the fractal dimension. However, the fractal dimension is not a unique descriptor, as it was shown that
two structures may have the same fractal dimension but are fundamentally
different \cite{Hayakawa1987}. In order to describe structures more deeply and unequivocally, the determination of the multifractal properties of an associated measure (e.g. growth probability) offers a suitable supplement to the sole measurement of their fractal dimension. This is an entropy-based approach \cite{Kinsner2005}, classified this way to differentiate it from other analyses that rely only on metric concepts. Here the scaling properties are analysed for variations in different parts of the pattern, which are overlooked by a simple measurement of the fractal dimension. As it is arduous to treat growth models with nonlocal rules analytically \cite{Barabasi1995}, it may be interesting to characterise an associated measure such as the growth probability to gain some insight about the process.

A way to describe the multifractal behaviour is through the generalised
dimensions $D_{q}$ (also known as R\'enyi dimensions). If one covers
the support of the measure (set of all points where the measure is
positive) with a set of boxes of size $l$ and defines a probability
$P_{i}(l)$ (integrated measure) in the $i$th box, the generalised
dimensions $D_{q}$ correspond to the scaling exponents for the $q$th
moments of $P_{i}$, defined by $\sum\limits _{i}P_{i}^{q}(l)\sim l^{(q-1)D_{q}}$.
In this context, $q$ is typically referred to as the order $q$ of the generalised dimension $D_{q}$. Solving for $D_{q}$ and taking the limit of $l\rightarrow0,$ the
conventional expression for the generalised dimensions is given by
\begin{equation*}
D_{q}=\frac{1}{(q-1)}\underset{l\rightarrow0}{lim}\frac{\ln\underset{i}{\sum}P_{i}^{q}(l)}{\ln l}.
\end{equation*}
For the case $q=1$, the L'Hôpital's rule must be used; thus, 
\begin{equation*}
D_{1}=\underset{l\rightarrow0}{lim}\frac{\underset{i}{\sum}P_{i}(l)\log P_{i}(l)}{\ln l}.
\end{equation*}
The generalised dimensions are exponents that characterise the non-uniformity
of the measure; the positive orders $q$ accentuate the regions with higher
probabilities while the negative $q$'s the opposite. 

In early works, $D_{q}$ was only defined for $q\geq0$ \cite{Hentschel1983}, so the previous definition had the particularity that the first value of the set, i.e. $D_{q=0}$, corresponded to the Hausdorff dimension of the support (because all the boxes have the same weight). There are also specific names for other certain values. For example, $D_{q=1}$ is known as the information
dimension, which is interesting in the case of diffusion-limited aggregations,
since it can be physically interpreted as the fractal dimension of
the active region, i.e., the unscreened region \cite{Hayakawa1987};
$D_{q=2}$ is known as the correlation dimension; $D_{q\rightarrow\text{\textpm}\text{\ensuremath{\infty}}}$
are known as the Chebyshev dimensions, which are calculated with the
maximum and minimum probabilities, respectively; equivalences with
other dimensions definitions can be made, even for fractional $q$
values \cite{Kinsner2005}.

This is not the only multifractal analysis possible. In other works, the singularity spectrum is computed \cite{Halsey1986, Chhabra1989a}, which is closely related to the Rényi dimensions by a Legendre transform. Temporal fractals can also be studied, where the local scaling properties are now related to time behaviour \cite{Ivanov2009}. We will use the generalised dimensions to characterise quantitatively the patterns produced by our model.

\section{Model}

In this paper, we model the growth of non-motile bacterial colonies
under different environmental conditions, specifically, nutrient concentration
and nutrient diffusion. The growth rules are inspired by biology,
as we capture the essential characteristics of bacteria without losing
simplicity. We consider a two-dimensional and off-lattice space, 
which allows us to consider mechanical interactions between the agents, as we will explain below.

There are two kinds of particles in the model, nutrient particles and bacterial
cells. Both of them have physical properties such as size, mass, position,
velocity and might have applied forces. Nutrient particles are idealised
as Brownian particles, so its initial velocities follow the Maxwell-Boltzmann
distribution and evolve according to a Langevin equation of the form
$m\dot{v}(t)=-\frac{\kappa T}{D}v(t)+f(t),$ where $\kappa$ is
the Boltzmann's constant, $T$ is the temperature and $D$ is the
diffusion coefficient. The function $f(t)$ is a stochastic force
whose components follow a Gaussian probability distribution with mean
zero and standard deviation $\sigma=\kappa T\sqrt{2/D}$. The Langevin
equation is numerically integrated using a small time step $\varDelta t$,
following the explanations in \textsl{The Fokker-Planck Equation}
by H. Risken \cite{Risken1989}. Nutrient particles are considered point-like,
non-interacting with each other and with a small mass. The bacterial
cells are modelled as rigid circles with radius \textit{$r_{b}$},
so they can interact with each other through normal forces. The numerical
values used in the simulations can be found in Table \ref{tab:Numerical-values-used}.

\begin{table}
\caption{Numerical values used for the simulation.\label{tab:Numerical-values-used} }
\begin{centering}
\begin{tabular}{lcr}
\hline 
Variable & Value {[}arbitrary units{]}\tabularnewline
\hline 
$kT$ & $1$\tabularnewline
$\Delta t$ & $0.01$\tabularnewline
$r_{b}$ & $1$\tabularnewline
$m$ (nutrient mass) & $0.0001$\tabularnewline
\hline 
\end{tabular}
\par\end{centering}
\end{table}

\begin{figure}[h]
\begin{centering}
\includegraphics[width=5cm]{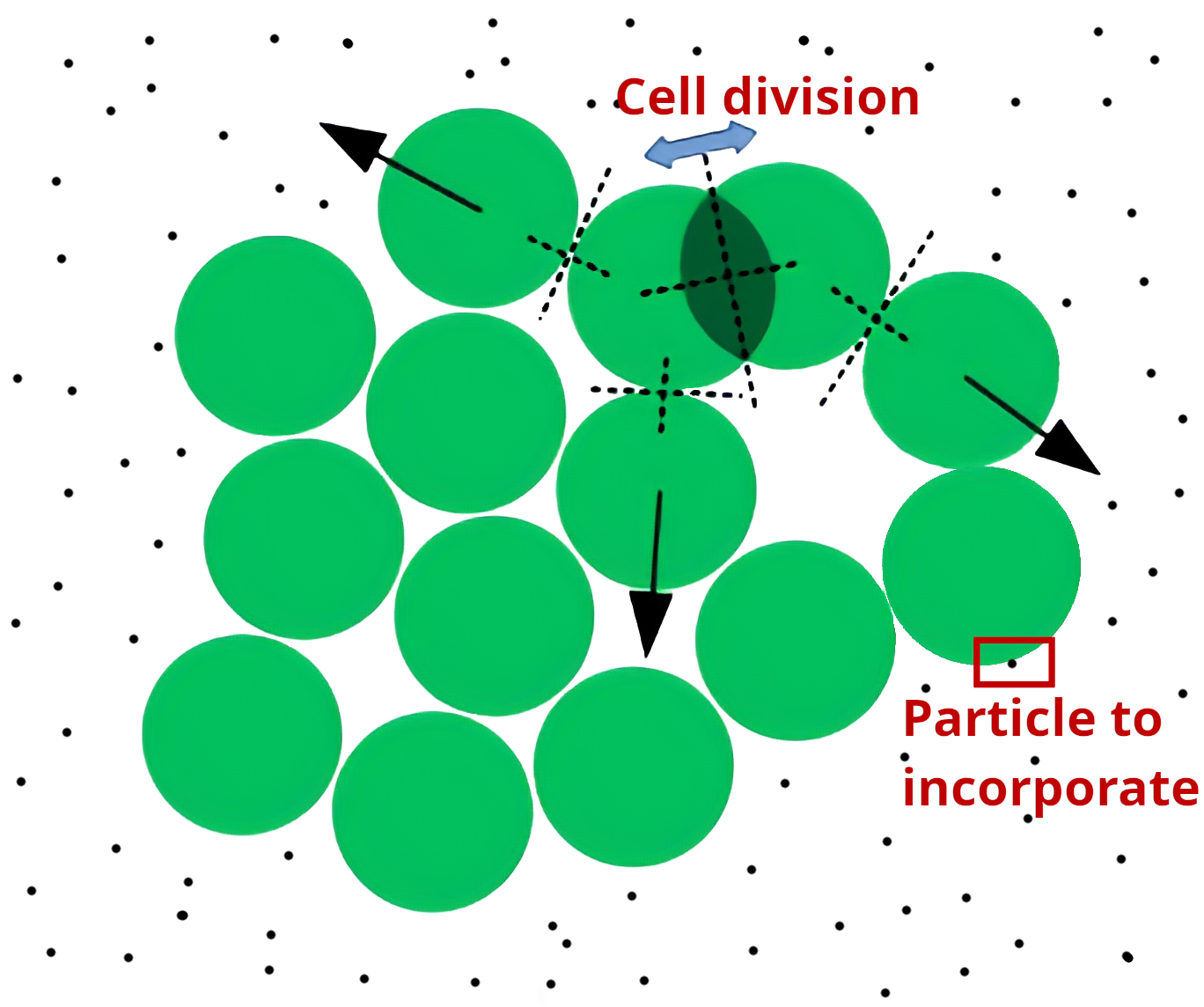}
\par\end{centering}
\caption{Feeding and reproduction. Bacteria cells (circles) can absorb the particles diffusing in the medium (dots) and duplicate, causing collisions with neighbours, which
move in the direction pointed by the arrows. The double blue arrow
(colour online) points the direction in which the newborn bacteria
disaggregate. \label{fig:Bacteria-cells-can}}
\end{figure}

In addition to the physical properties, bacterial cells have two biological
characteristics: they can be fed and reproduce. The first one is an
interaction with the nutrients, which are absorbed by the cell
when they are in contact. Reproduction is the process by which the
bacterium duplicates: an identical copy of the cell is generated in
the same position as the original, so they overlap. Then, they are
disaggregated by opposing velocities in a random direction.
As a consequence, these cells may collide elastically with neighbours,
according to the mechanics of rigid bodies, as shown in Fig. \ref{fig:Bacteria-cells-can}.
When the cells stop overlapping, they stop moving and become static because
the medium is considered to be very viscous so that the momentum gained by
the collisions is immediately dissipated. All the calculations to resolve collisions and overlaps are based on an iterative constraint solver introduced by Erin Catto \cite{Catto2005}. 

In summary, reproduction causes both the movement of newborn cells and their neighbours, representing sliding motility.

The simulation begins with a single cell in the origin of coordinates in
a 2D substrate and a given quantity of nutrient particles diffusing
in space, according to the concentration and diffusion coefficient
specified. When a nutrient particle touches the cell, it is absorbed
and the bacteria duplicates. Now the colony is formed by two cells,
which can absorb nutrients and reproduce. The process continues in this way and
the colony grows progressively. All the bacteria have a time delay
(20 integration time steps), during which they can't duplicate; this
rule ensures that no duplication occurs while newborn cells are still
overlapping and it is consistent with biological observations,
e.g., \textit{Bacillus subtilis} species has a delay of $\sim 25$ minutes between duplications \cite{Golding1998}.

\begin{figure}
\begin{centering}
\includegraphics[width=5cm]{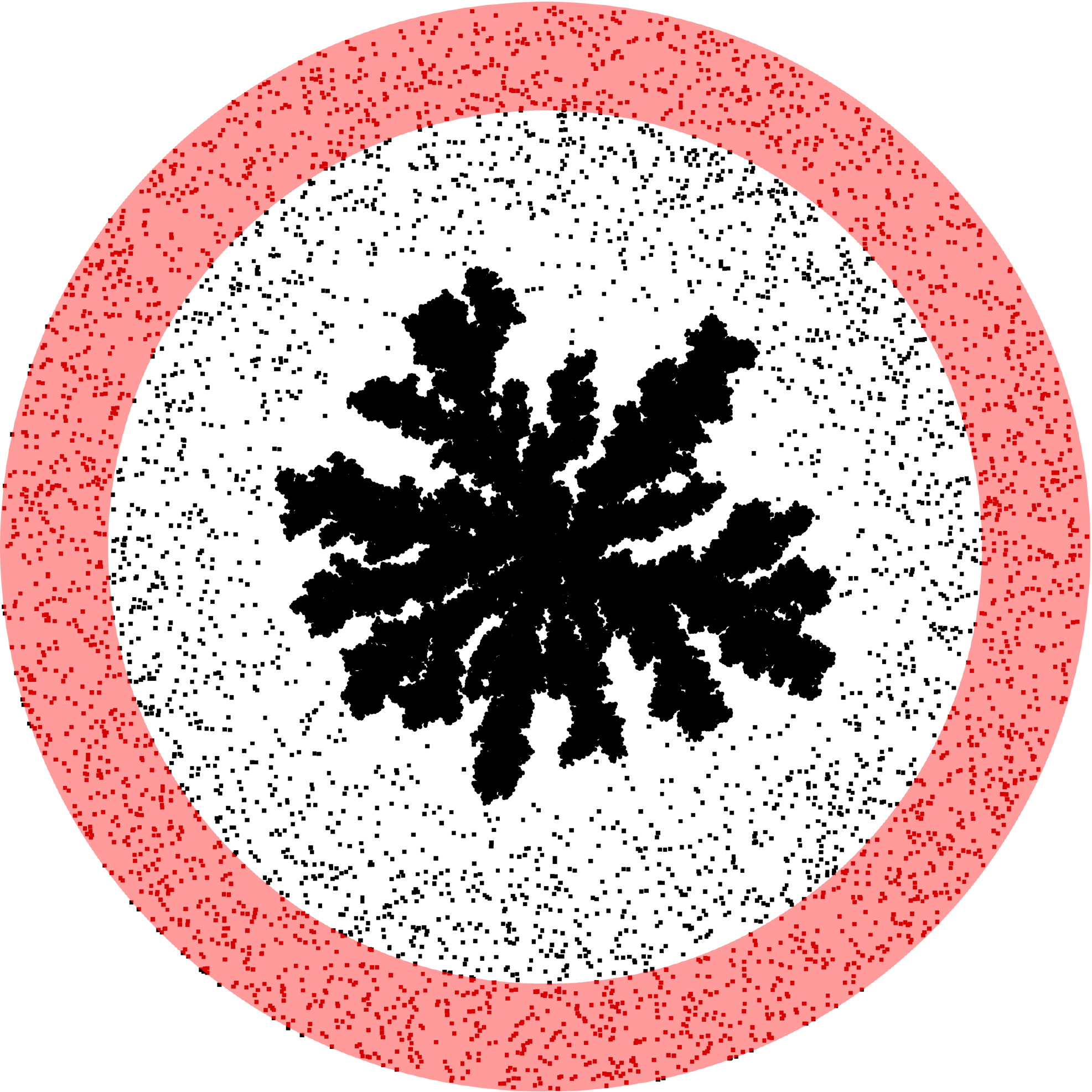}
\par\end{centering}
\caption{A ring surrounding the colony (in red) acts as a nutrient reservoir. The concentration
of nutrient is constant inside it, but decreases in the proximity
of the colony, because of the feeding. The distance between the ring
and the colony is \textit{$60r_{b}$} at least. \label{fig:food-ring}}

\end{figure}

In order to keep the nutrient concentration constant, there is a ring that acts as a nutrient reservoir located at a given distance from the most external position of the bacteria. This distance increases progressively as the colony grows, so the separation
is \textit{$60 r_{b}$} at least. The nutrient concentration within the ring ($r>60 r_{b}$) is kept constant at a specified value. But, closer to the colony ($r<60 r_{b}$), the concentration drops due to nutrient absorption by the bacteria. Periodic boundaries conditions are considered for the outer side of the ring (Fig. \ref{fig:food-ring}).

The growth stops when the colony reaches a radius of \textit{$600 r_{b}$},
when characteristic patterns are fully developed. This implies
that we have up to half a million cells forming the colony, depending
on the parameters.

\section{Results and discussions}

\subsection{First characterization of the structures}

\begin{figure*}% >>>
  \captionsetup[subfloat]{farskip=4pt}
  \begin{centering}
  \begin{minipage}[t]{.85\linewidth}
    \vspace{0pt}
    \subfloat{
      \stackinset{l}{0.12\linewidth}{t}{0.01\linewidth}{(a) \label{fig:muestra}}{
      \includegraphics[width=0.948\linewidth]{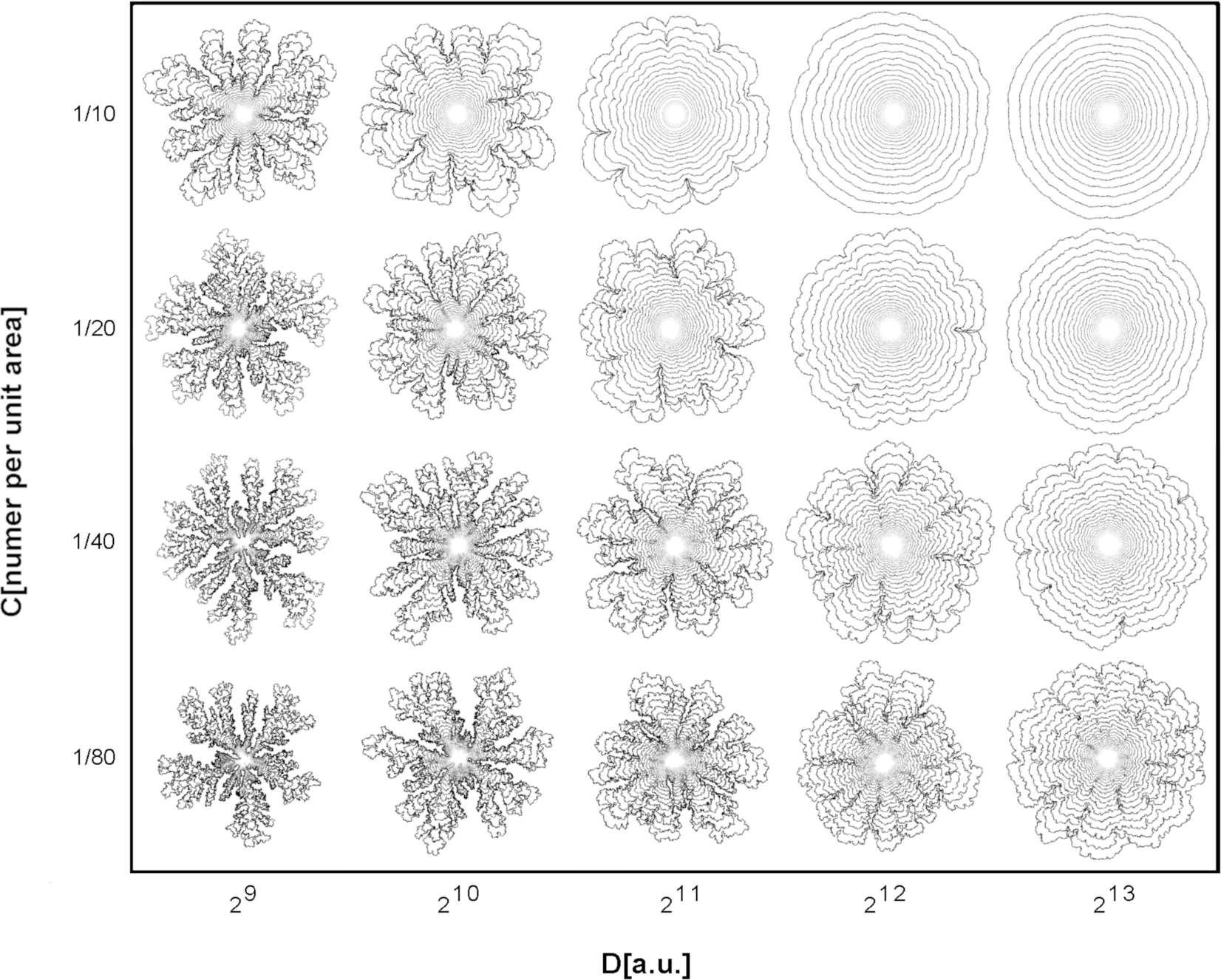}}}%
  \end{minipage}%
  \hfill
  \begin{minipage}[t]{.15\linewidth}
    \vspace{0pt}
    \subfloat{
      \stackinset{l}{0.1\linewidth}{t}{0.05\linewidth}{\textcolor{white}{(b) \label{fig:experiments}}}{
      \includegraphics[width=\linewidth]{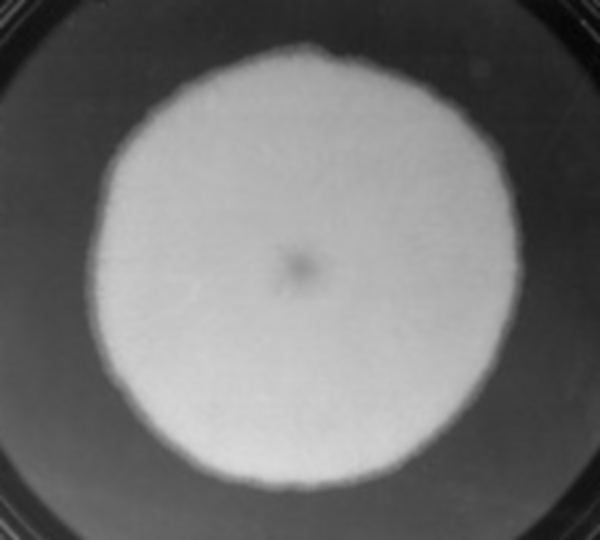}}}\\
    \subfloat{
      \stackinset{l}{0.1\linewidth}{t}{0.05\linewidth}{\textcolor{white}{(c)}}{
      \includegraphics[width=\linewidth]{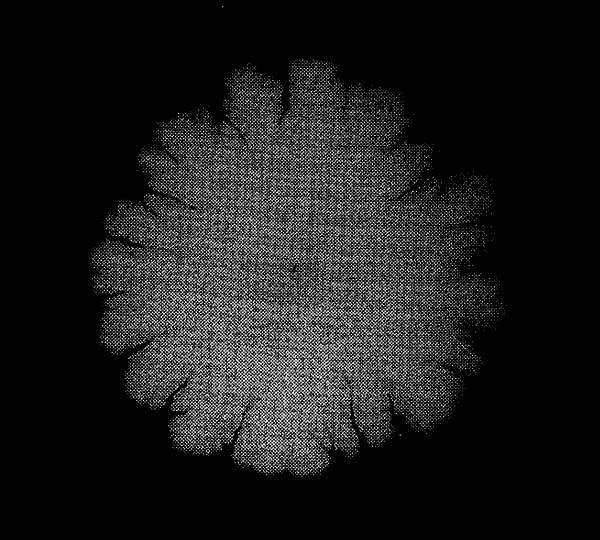}}}\\
    \subfloat{
      \stackinset{l}{0.1\linewidth}{t}{0.05\linewidth}{\textcolor{white}{(d)}}{
      \includegraphics[width=\linewidth]{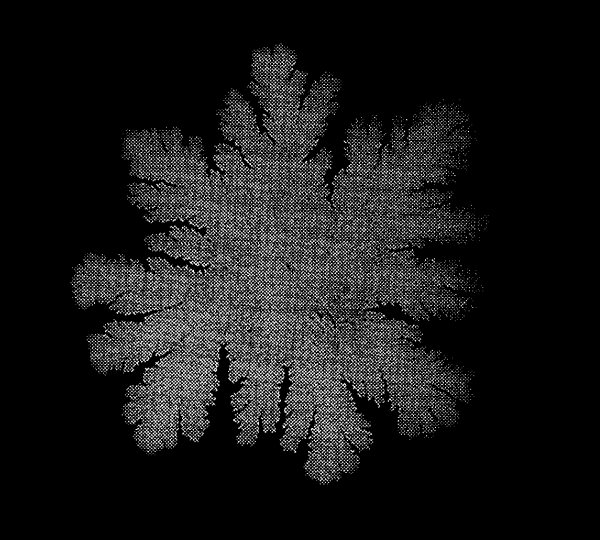}}}\\
    \subfloat{
      \stackinset{l}{0.1\linewidth}{t}{0.05\linewidth}{\textcolor{white}{(e)}}{
      \includegraphics[width=\linewidth]{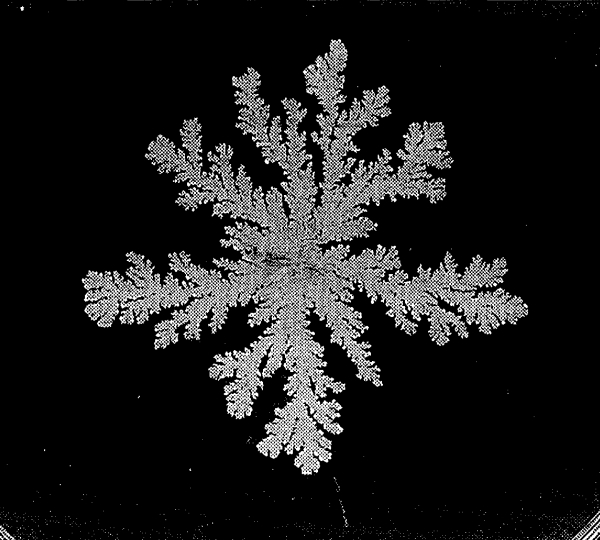}}}\\
  \end{minipage}%
  \caption
    {%
      Examples of different morphologies. In (a) there is a sample of the model predictions. On the $X$-axis, the diffusion coefficient $D$ is varied (in arbitrary units), while on the $Y$-axis the nutrient concentration is varied (measured as the number of particles per unit area). Each curve, in a different shade of grey, corresponds to a different time. All other sub-figures correspond to experiments. The parameter $C_{a}$ corresponds to the agar concentration, which determines the hardness of the substrate, and $C_{n}$ corresponds to the nutrient concentration. (b) $C_{a}=10g/l;C_{n}=20g/l$. (c) $C_{a}=9g/l;C_{n}=4.5g/l$. (d) $C_{a}=8g/l;C_{n}=3g/l$. (e) $C_{a}=9g/l;C_{n}=1g/l$. Figure (b) is from \cite{Matsushita1998}, reprinted with permission from Elsevier. Figures (c), (d) and (e) are from \cite{Ohgiwari1992}, \textcopyright(1992) The Physical Society of Japan, reproduced with permission.%
    }%
  \end{centering}
\end{figure*}

In order to see the variety of morphologies that the model can produce,
we choose several different values of nutrient concentration and nutrient
diffusion, and register the position of the bacteria along the perimeter
of the colony over time and average over a hundred realizations for
each set of parameters. A morphology diagram is shown in Fig. \ref{fig:muestra},
where it can be seen that it is possible to generate round and compact
colonies, as well as ramified, going through a variety of intermediate
patterns. Similar morphological crossover can be seen in Figs. \ref{fig:experiments}--e,
which corresponds to the experiments carried out in \cite{Matsushita1998, Ohgiwari1992}.
The fractal dimension of intermediate patterns in experiments was
reported for the case of the most ramified one. In Table \ref{tab:dimension-fractal},
we summarise the results found in the bibliography \cite{Matsushita1990,Ohgiwari1992}
and ours (for the most ramified cases), which are in good agreement.

% <<<

\begin{table}
\caption{Fractal dimension $D_{f}$. Values reported in experimental works, our model
and DLA (the error is the standard deviation). Only the results for the most branched cases are included with $C=1/80, D=512$ and $C=1/40, D=512$. Greater values than for the case of DLA are expected, where only one particle diffuses at a time, unlike our case where we have many ($C>0$). \label{tab:dimension-fractal}}

\centering{}%
\begin{tabular}{lcr}
\hline 
 & $D_{f}$\tabularnewline
\hline 
Model {[}$C=1/80;D=512${]} & $1.760\pm0.004$\tabularnewline
Model {[}$C=1/40;D=512${]} & $1.778\pm0.003$\tabularnewline
Experiment \cite{Matsushita1990} & $1.73\pm0.02$\tabularnewline
Experiment \cite{Ohgiwari1992} & $1.70\pm0.02$\tabularnewline
DLA \cite{Hayakawa1987, Barabasi1995} & $1.71\pm0.01$\tabularnewline
\hline 
\end{tabular}
\end{table}

\subsection{Scaling properties}

\begin{figure}
\begin{centering}
\subfloat{\begin{centering} \label{fig:SvsNnocoll}
\stackinset{l}{0.7in}{t}{0.2in}{(a)}{
\includegraphics[width=6.5cm]{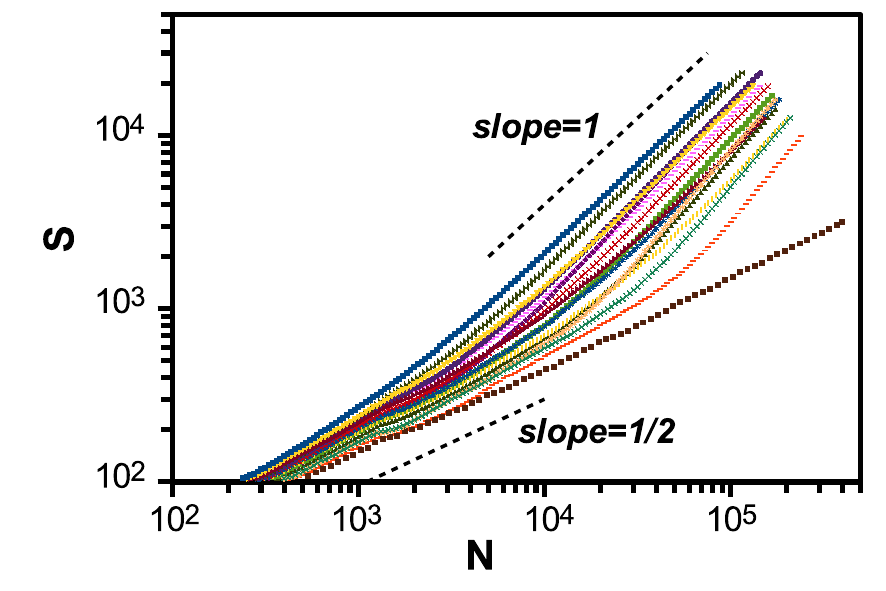}}
\par\end{centering}
}
\par\end{centering}
\begin{centering}
\subfloat{\begin{centering} \label{fig:collap}
\stackinset{l}{0.7in}{t}{0.2in}{(b)}{
\includegraphics[width=6.5cm]{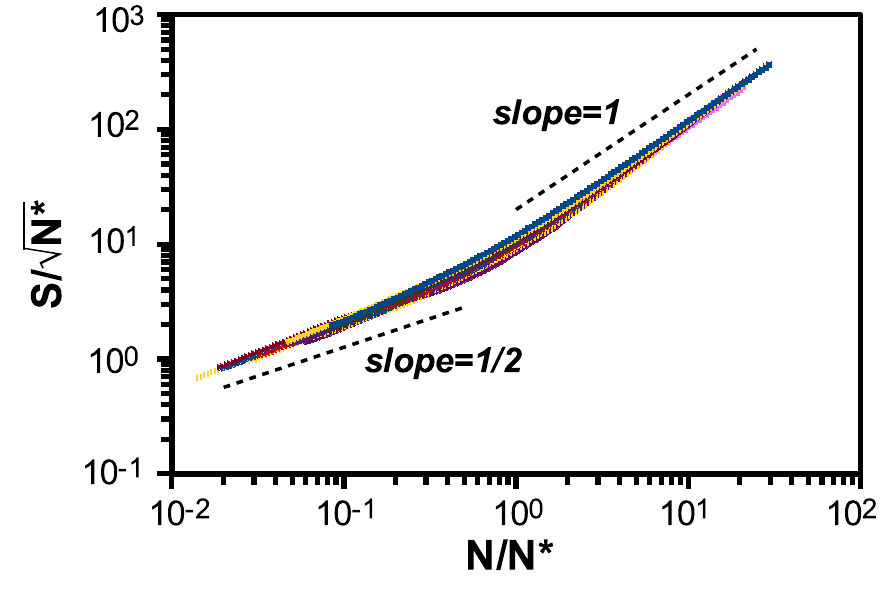}}
\par\end{centering}
}
\par\end{centering}
\caption{Power-law regimes. (a) Results, averaging over all realizations, of the number of bacteria
in the interface $S$ versus the total number of cells $N$. (b) Collapsed curves of $S$ versus $N$. Both plots are double-logarithmic.\label{fig:a)-S-vs}}
\end{figure}

Despite using different values for the parameters, it is observed
that the curves of the number of bacteria at the interface S versus
the total number N show two power-law regimes. The first regime corresponds
to initial compact structures $S\sim N^{1/2}$, while the second regime
corresponds to ramified structures with $S\sim N$, as shown in Fig.
\ref{fig:SvsNnocoll}. These two behaviours are characteristic of
the Eden and DLA models, respectively.

To characterise the crossover between regimes, we compute the total number of bacteria $N^{*}$ at which the crossover takes place. To achieve this, we simply fit power-law functions in the tails of the $S$ vs $N$ curves and compute the intersection. More sofisticated methods for an automatic determination of crossovers can be found in \cite{Bashan2008}. After dividing
$N$ by $N^{*}$ in each of the data sets, the $Y$-axis is divided
by some value $S^{*}$ looking for a satisfactory collapse of the curves.
We found that the best collapse occurs when $N^{*}=S^{*2}$, as shown
in Fig. \ref{fig:collap}.

\begin{figure}
\begin{centering}
\subfloat{\begin{centering}
\stackinset{l}{0.8in}{t}{0.2in}{(a)}{
\includegraphics[width=6.5cm]{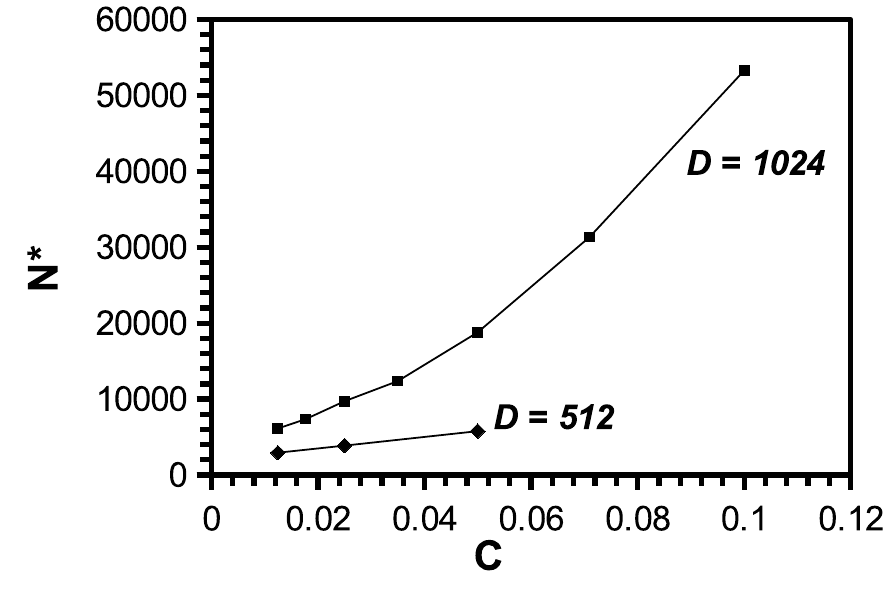}}
\par\end{centering}
}
\par\end{centering}
\begin{centering}
\subfloat{\begin{centering}
\stackinset{l}{0.8in}{t}{0.2in}{(b)}{
\includegraphics[width=6.5cm]{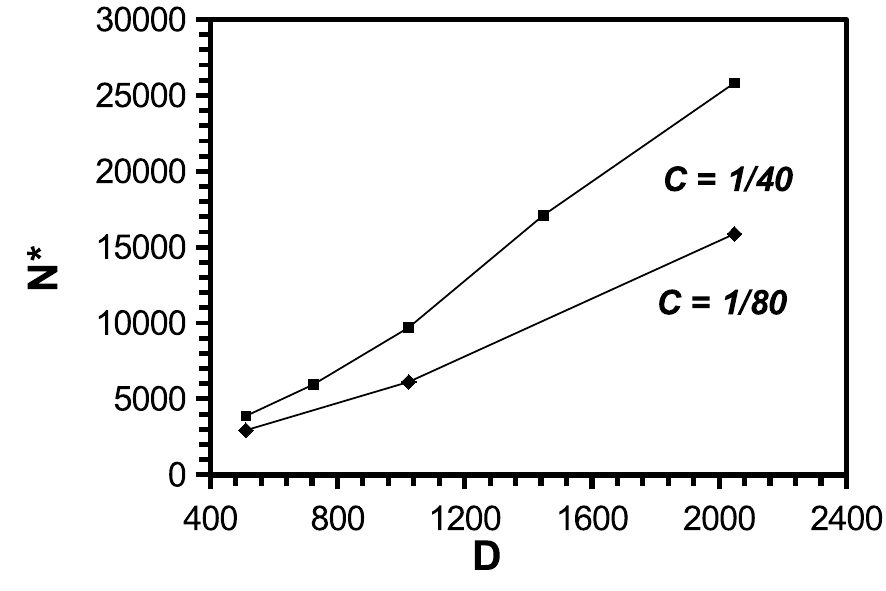}}
\par\end{centering}
}
\par\end{centering}
\caption{Relationship between $N^{*}$ and the parameters. (a) $N^{*}$ is
plotted against the nutrient concentration $C$, leaving the diffusion
coefficient $D$ constant; (b) the same but for $D$. \label{fig:N*vs-D-and}}

\end{figure}

It can be seen that $N^{*}$ depends on the diffusion $D$ and the concentration
$C$, having an increasing relationship with both (Fig. \ref{fig:N*vs-D-and}).

Taking these observations into account, an attempt is made to establish
a scaling law. We know that the behaviour of S is:

\begin{equation}
S\sim\begin{cases}
N^{1/2}, & N\ll N^{*}\\
N, & N\gg N^{*}
\end{cases},
\end{equation}

where $N^{*}=N^{*}(C,D)$. The curves collapse dividing $N$ and $S$
by $N^{*}$ and $S^{*}$, respectively, so:

\begin{equation}
S/S^{*}\sim\begin{cases}
(N/N^{*})^{1/2}, & N/N^{*}\ll1\\
N/N^{*}, & N/N^{*}\gg1
\end{cases}.
\end{equation}

Then, having validated the relationship $N^{*}=S^{*2}$ and proposing
the scaling function 

\begin{equation}
f(x)\sim\begin{cases}
const, & x\ll1\\
x, & x\gg1
\end{cases},
\end{equation}

the relation between $N$ and $S$ can be written as:

\begin{equation}
S=N^{1/2}f[(\frac{N}{N^{*}})^{1/2}].
\end{equation}

This result suggests that if we allow $N$ to grow sufficiently, branches will always
be generated, after a critical number of cells is reached, dependent
on the parameters $C$ and $D$.

\subsection{Multifractality of the growth probability}

The growth probability of each region of the colony can give information
about why a certain pattern displays. Every cell duplicates when a
nutrient particle is captured, so the growth probability is associated
with the probability that a diffusing particle reaches the site where
the cell is. We use two methods to estimate this probability, focusing
on the final stage of the colony. The first one consists on counting
how many nutrient particles are absorbed by each cell without letting
it duplicate, i.e., the colony is \textquotedbl frozen\textquotedbl{}
and the growth probability of each cell is computed dividing this
counting by the total of particles incorporated by the whole colony
(we use approximately $10^{5}$ particles). We will refer to this
method as C.M. The disadvantage with this method is that it does not
estimate low probabilities well, because several million particles
may be captured by the colony in total, but the internal regions may
hardly incorporate any. Due to this, we also solve the Laplace equation
$\nabla^{2}\phi=0$, where $\phi$ represents the nutrient concentration,
by the relaxation method \cite{Li2003}, where $\phi=1$ at infinity
and $\phi=0$ along the perimeter of the colony, as it can be seen
that the growth probability is proportional to the gradient of the
potential $\nabla\phi$ \cite{Hayakawa1987}. We use an iteration error
of $10^{-5}$, after checking that the multifractal curves do not
vary appreciably. In order to use this method, referred to as L.M. henceforth, properly, space has to be discretised, so some differences with the C.M. are expected. 

\begin{figure}
\centering{}\subfloat[Absorption $>1$]{\centering{}\includegraphics[width=2.9cm]{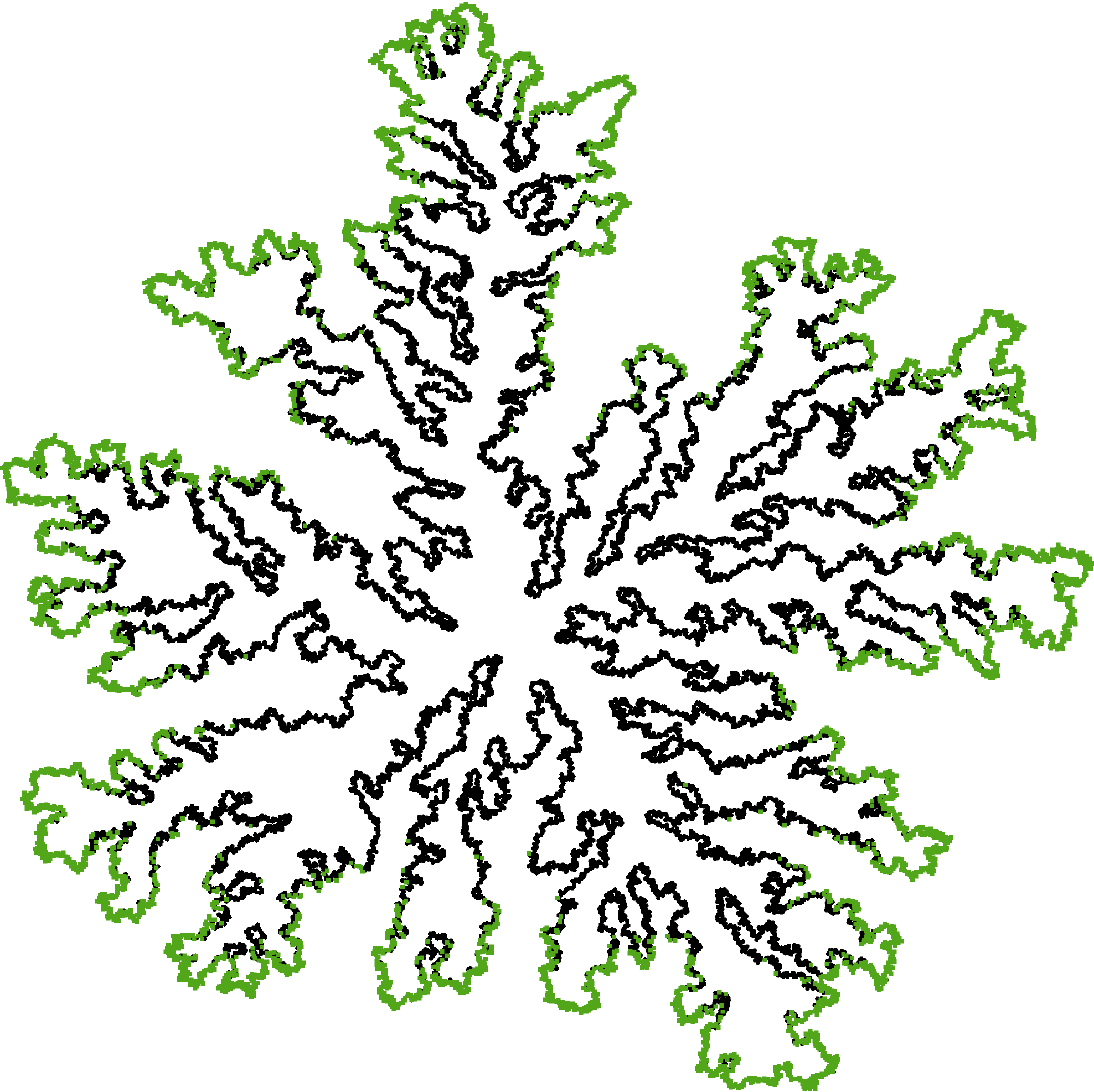}}
\subfloat[Absorption $>20$]{
\begin{centering}
\includegraphics[width=2.9cm]{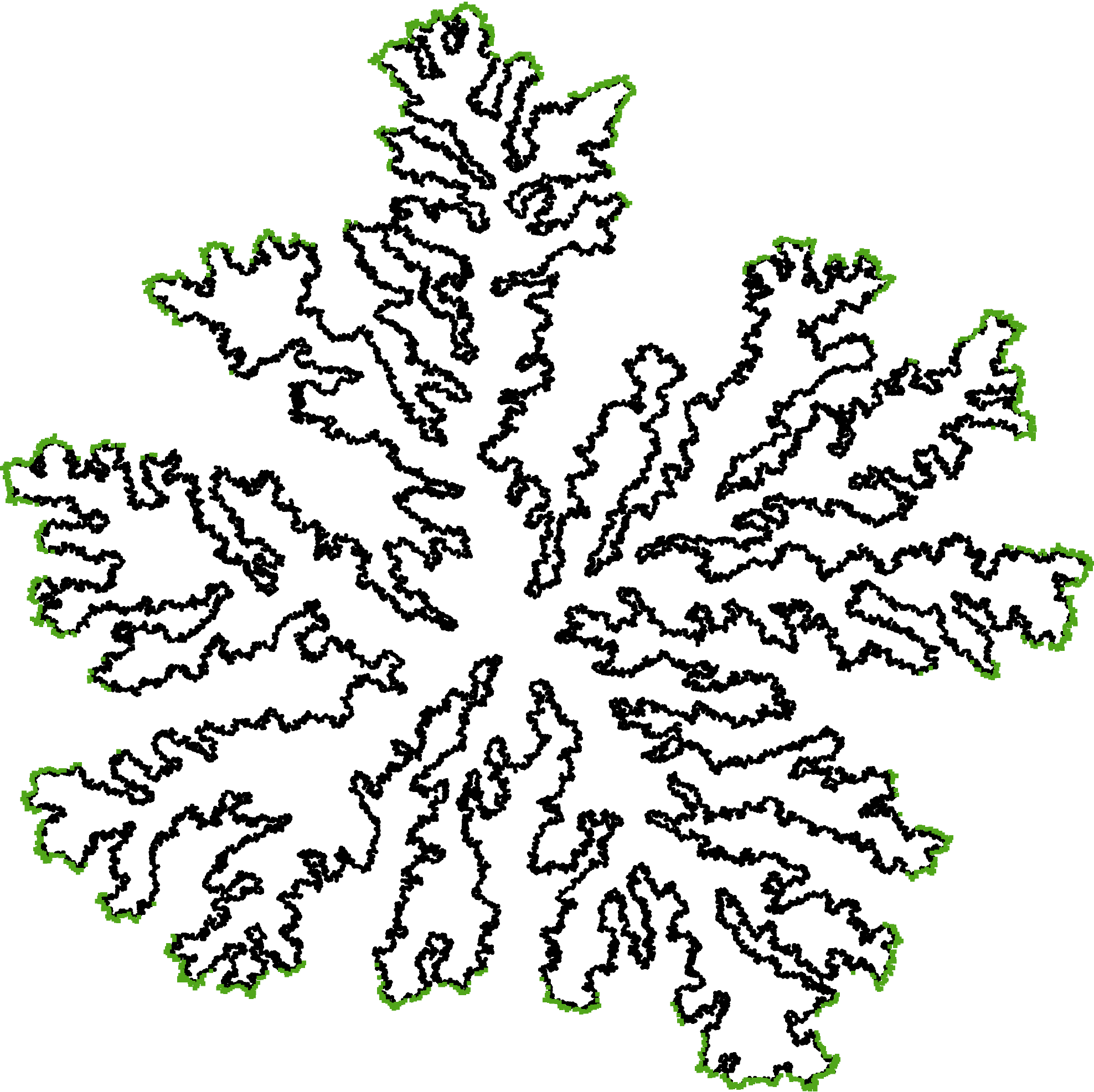}
\par\end{centering}
}

\subfloat[Absorption $>80$]{\begin{centering}
\includegraphics[width=2.9cm]{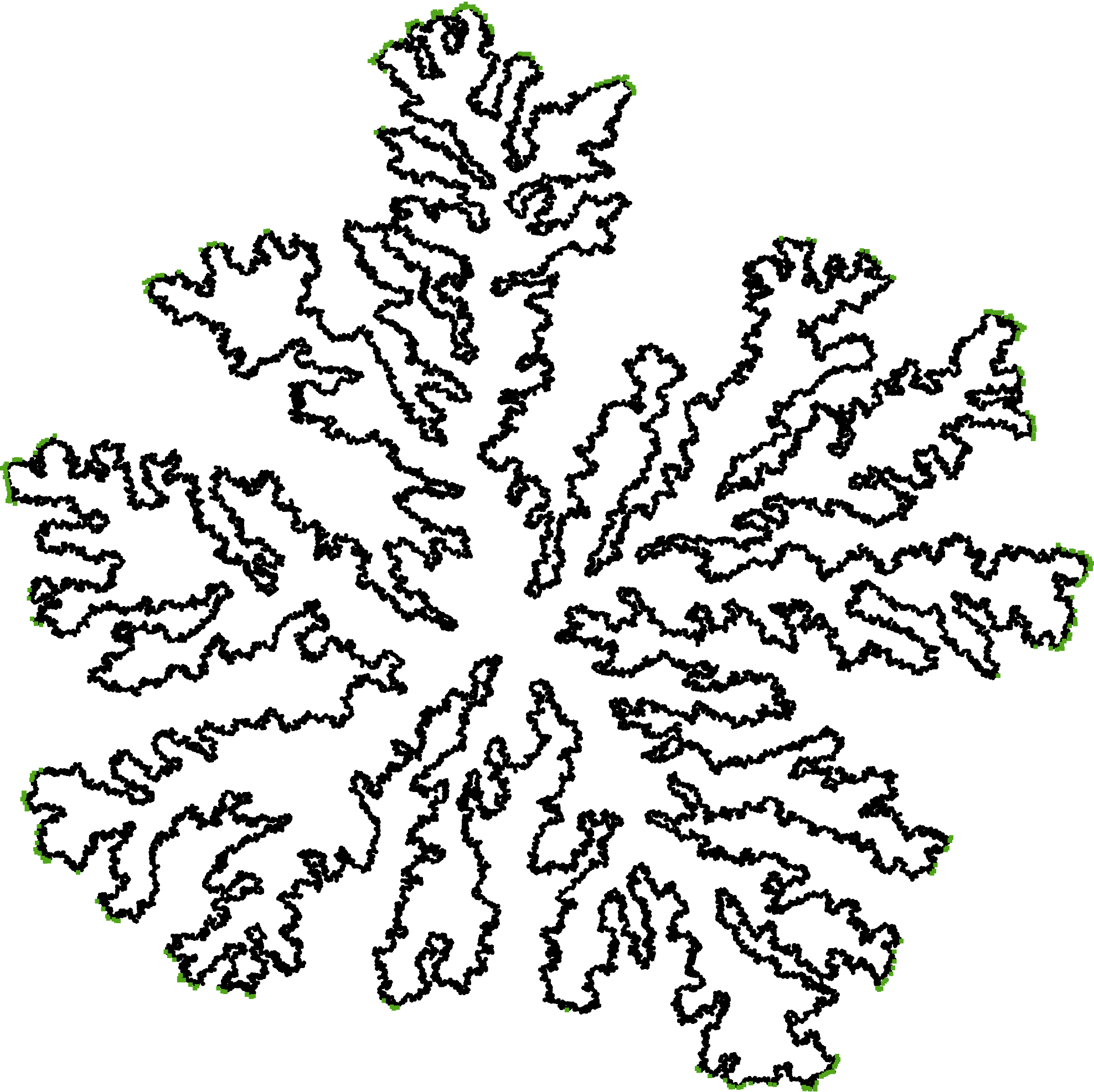}
\par\end{centering}
}
\caption{Example of the C.M. results for a branched colony. After $10^{5}$
particles are captured, the green colour marks the cells who absorbed
more than (a) 1 particle, (b) 20 particles and (c) 80 particles. \label{fig:active-region}}
\end{figure}

In Fig. \ref{fig:active-region}, it is shown how uneven is the number
of nutrient particles consumed between the outer and inner regions of
a ramified colony. This phenomenon is usually referred to as `shadowing'
or `screening' effect and is more or less noticeable depending
on the parameters. As the structures that emerge from the simulations
are fractal, the proper way to study this effect is by the multifractal
formalism, explained in the first section. 

\begin{figure}
\begin{centering}
\subfloat{\begin{centering}
\stackinset{l}{0.7in}{b}{0.45in}{(a)}{
\includegraphics[width=6.5cm]{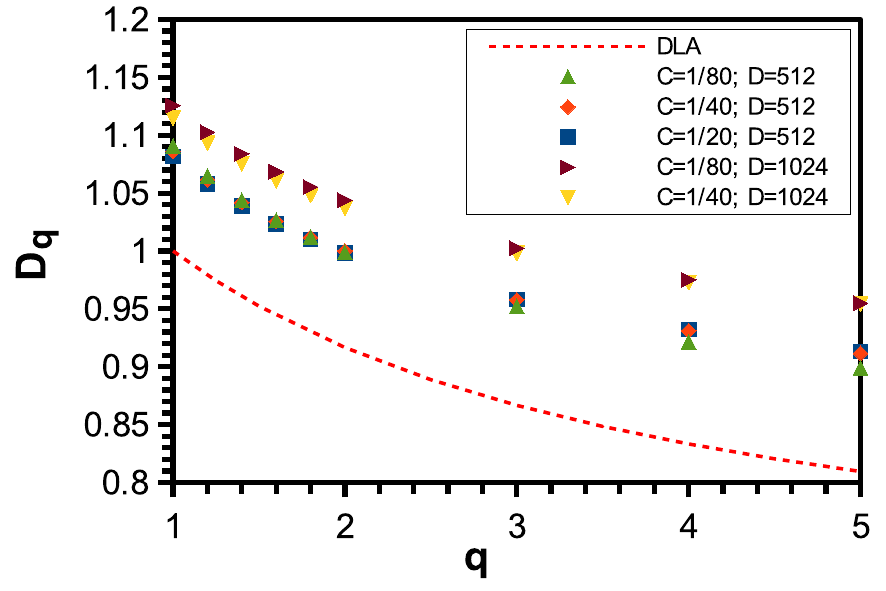}}
\par\end{centering}
}
\par\end{centering}
\begin{centering}
\subfloat{\begin{centering}
\stackinset{l}{0.7in}{b}{0.45in}{(b)}{
\includegraphics[width=6.5cm]{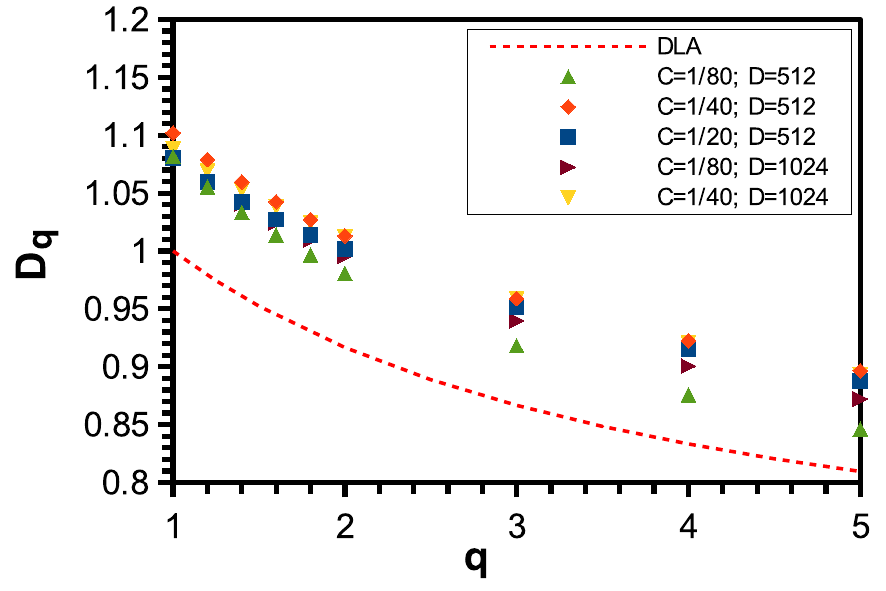}}
\par\end{centering}
}
\par\end{centering}
\caption{Generalised dimension curve for $q>1$. Different icons correspond
to different values of nutrient concentration $C$ and diffusion coefficient
$D$. Red dashed line corresponds to DLA (colour online). Results using
the (a) C.M. and (b) L.M.\label{fig:Dq>1}}
\end{figure}

In Fig. \ref{fig:Dq>1}, the generalised dimension $D_{q>1}$ curve
is plotted for different morphologies. It can be seen that the probability
associated with a ramified colony presents a strong multifractality
since $D_{q}$ varies significantly with $q$. It is also included
in Fig. \ref{fig:Dq>1} the curve for a diffusion-limited aggregate
\cite{Matsushita1987} for comparison. The standard deviations are
presented in Table \ref{tab:puntos-caracteristicos}. Unfortunately,
this analysis cannot be carried out in very compact colonies since
the fractal regime is very short to be reliable or it is not observable. Nor can it be done in the early stages of the growth process for the same reason.

\begin{figure}
\begin{centering}
\includegraphics[width=6.5cm]{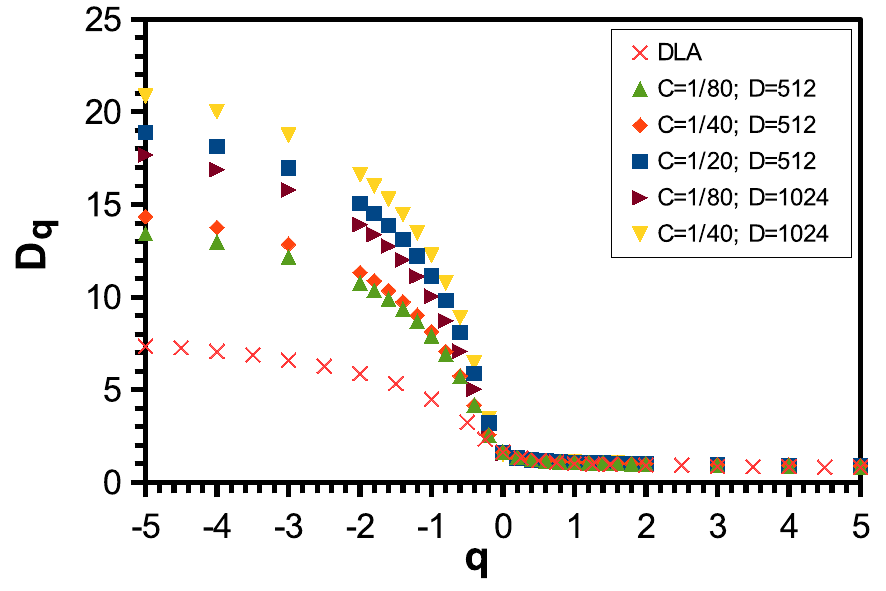}
\par\end{centering}
\caption{Generalised dimension curve including negative values of $q$. Only
the L.M. results are plotted. Different icons correspond to different
values of nutrient concentration $C$ and diffusion coefficient $D$. Note that the curves do not collapse for $q>0$, the difference between them cannot be seen on this scale (see Fig. \ref{fig:Dq>1}).\label{fig:Dq-y-f(a)}}
\end{figure}

In Fig. \ref{fig:Dq-y-f(a)}, the generalised dimension is plotted
again, but now including the $q<1$ interval. Only the results obtained
by the L.M. can be used in this interval. The value of $D_{q\ll-1}$
is very interesting, because it quantifies the shadowing effect, making
evident the differences between different morphologies. It is worth
noting that $D_{q=0}$ should not be equal to the fractal dimension
$D_{f}$ presented in the previous subsection because $D_{q=0}$
is the fractal dimension of the support of the measure, i.e., the
perimeter of the colony, while $D_{f}$ is the fractal dimension of
the area. All of these characteristic points are summarised in Table
\ref{tab:puntos-caracteristicos}, which also includes the values
corresponding to diffusion-limited aggregation as a comparison \cite{Hayakawa1987,Matsushita1987}.
The asymptotic values are estimated with $q=25$ and $q=-25$.

\begin{table*}
\caption{Some characteristic values of the generalised dimension $D_{q}$ (the
error is the standard deviation). DLA results found in \cite{Hayakawa1987,Matsushita1987}
are included for comparison. \label{tab:puntos-caracteristicos}}

\begin{centering}
\begin{tabular}{ccccccc}
\toprule 
\multirow{2}{*}{} & \multicolumn{2}{c}{$C=1/20;D=512$} & \multicolumn{2}{c}{$C=1/40;D=512$} & \multicolumn{2}{c}{$C=1/40;D=1024$}\tabularnewline
\cmidrule{2-7} 
 & C.M. & L.M. & C.M. & L.M. & C.M. & L.M.\tabularnewline
\midrule
 $D_{q=0}$ & $-$ & $1.60\pm0.01$ & $-$ & $1.63\pm0.01$ & $-$ & $1.53\pm0.01$\tabularnewline
 $D_{q=1}$ & $1.08\pm0.01$ & $1.08\pm0.05$ & $1.09\pm0.01$ & $1.10\pm0.09$ & $1.11\pm0.01$ & $1.09\pm0.08$\tabularnewline
 $D_{q=2}$ & $1.00\pm0.02$ & $1.00\pm0.09$ & $1.00\pm0.02$ & $1.01\pm0.12$ & $1.04\pm0.01$ & $1.01\pm0.13$\tabularnewline
 $D_{q\gg1}$ & $0.82\pm0.04$ & $0.75\pm0.09$ & $0.82\pm0.04$ & $0.79\pm0.15$ & $0.87\pm0.04$ & $0.78\pm0.15$\tabularnewline
 $D_{q\ll-1}$ & $-$ & $19.0\pm1.5$ & $-$ & $15.5\pm1.5$ & $-$ & $18.7\pm2.1$\tabularnewline
\bottomrule
\end{tabular}
\par\end{centering}
\centering{}%
\begin{tabular}{ccccccc}
\toprule 
\multirow{2}{*}{} & \multicolumn{2}{c}{$C=1/80;D=512$} & \multicolumn{2}{c}{$C=1/80;D=1024$} & \multicolumn{2}{c}{DLA}\tabularnewline
\cmidrule{2-7} 
 & C.M. & L.M. & C.M. & L.M. & \cite{Hayakawa1987} & \cite{Matsushita1987}\tabularnewline
\midrule
 $D_{q=0}$ & $-$ & $1.64\pm0.01$ & $-$ & $1.56\pm0.01$ & $1.64\pm0.01$ & \tabularnewline
 $D_{q=1}$ & $1.09\pm0.01$ & $1.08\pm0.07$ & $1.13\pm0.01$ & $1.08\pm0.06$ & $1.04\pm0.01$ & $1$\tabularnewline
 $D_{q=2}$ & $1.00\pm0.02$ & $0.98\pm0.11$ & $1.04\pm0.01$ & $1.00\pm0.10$ & $-$ & $0.92$\tabularnewline
 $D_{q\gg1}$ & $0.80\pm0.03$ & $0.73\pm0.14$ & $0.85\pm0.03$ & $0.76\pm0.13$ & $0.67\pm0.03$ & $0.66$\tabularnewline
 $D_{q\ll-1}$ & $-$ & $14.0\pm1.4$ & $-$ & $18.3\pm1.9$ & $\simeq9$ & $-$\tabularnewline
\bottomrule
\end{tabular}
\end{table*}

Note that the curves of the generalised dimension are always above
the case of DLA. In the region of $q>0$, taking into account the
calculated standard deviations, the differences are not as noticeable,
but they are in the region of $q<0$. In this region, where the measurement
best distinguishes each case, they depart notoriously from the case
of DLA. Always considering branched cases, it is observed that higher
values are associated with higher $C$ and $D$ values, which can
be understood if we associate this measure with the screening phenomenon.
The larger are $C$ and $D$, the thicker and narrower the branches and the fjords become, respectively, so the screening increases to the interior areas.
Nevertheless, note that there is a limit on how large the parameters can be. If $C$ and $D$ are too large, so the branches disappear, the screening effect is barely noticeable. The growth probability becomes almost uniform, and the multifractality should be lost. 

\section{Conclusions}

The goal of this work is the construction of a model based on basic
theories of physics, capable of generating a variety of complex patterns
observed in bacteria colonies. Under the hypothesis that there is
a single collective movement mechanism behind the different morphologies
(sliding), the different results are achieved by varying parameters
of the environment outside the colony, without changing the behaviour
of the agents. Under these precepts, we manage to generate patterns
from the most round and compact to extremely ramified, going through
different intermediate morphologies. The different approaches used
to characterise the structures also allow comparison with the two
most studied models that predict patterns of bacterial colonies, the
Eden model and DLA.

The fractal dimension analysis of simulations with branched colonies
shows a fractal dimension compatible with
experiments \cite{Matsushita1990}. Although there are distinct differences between the present and the DLA model, their fractal dimensions are in good agreement, which suggests that the most ramified cases considered are close to the diffusion limit. 

On the other hand, the characterization using scaling laws show
that there are two characteristic growth regimes, one compact and
one branched. According to the relation found, the crossover between regimes occurs at a critical number of cells, that depends on the parameters for nutrition concentration $C$ and nutrition diffusion $D$. Thus, branches will always be generated for finite values of these parameters. 

The multifractal measurement shows strong multifractality for the ramified cases. In case the order $q$ of the generalised dimension $D_{q}$ is $q>0$, the curves of $D_{q}$ of the simulations of different values of $C$ and $D$ are close to each other, but in the region of $q<0$, their differences become notoriously. All these curves are, however, always above the DLA curve and, as expected, approach to the DLA curve for lower values for $C$ and $D$. Higher values are associated with higher $C$ and $D$ values because fjords into the interior are narrow.

Unfortunately, a more in-depth comparison between simulation and experimental data cannot be carried out due to the lack of quantitative experimental data. To date, there are almost exclusively qualitative characterizations of the morphologies of bacterial colonies, which only in some rare cases provide additional information on the fractal dimension (which is not a complete indicator). Although the methods that we use in this work to calculate the generalised dimensions cannot be used in experiments, there are other methods that might be used, such as the one described by Ohta and Honjo \cite{Ohta1988}, based on associating probabilities according to the variation of the area occupied by the colony in a certain section. Only such a deeper experimental analysis would offer a complete characterization of the processes involved in the structure formation of bacterial colonies and would allow contrasting the proposed model with experiments.

\section*{Conflict of interest}

Authors declare that there are no conflicts of interest.

\section*{Acknowledgements}

L.A.B. and L.V. thanks UNMdP and CONICET (PIP 00443/2014) for financial support. D.H. thanks UNMdP and CONICET (PIP 0100629/2013) for financial support.

\section*{Author contribution statement}
L.V. and D.H. designed the model. L.V. implemented the computational framework. All authors contributed to the analysis of the data and in the writing of the manuscript.

\bibliographystyle{epj.bst}
\bibliography{Citas}

\end{document}